# Security Challenges when Space Merges with Cyberspace

Vijay Varadharajan[1] and Neeraj Suri[2]

Space borne systems, such as communication satellites, sensory, surveillance, GPS and a multitude of other functionalities, form an integral part of global ICT cyber infrastructures. However, a focused discourse highlighting the distinctive threats landscape of these space borne assets is conspicuous by its absence. This position paper specifically considers the interplay of Space and Cyberspace to highlight security challenges that warrant dedicated attention in securing these complex infrastructures. The opinion piece additionally adds summary opinions on (a) emerging technology trends and (b) advocacy on technological and policy issues needed to support security responsiveness and mitigation.

## 1. Introduction

The escalation of security attacks against spaceborne assets are blurring the boundaries of space security and cyber security. With the increasing commercialization and militarization of the space sector, their attractiveness as a target and the consequent needs of cyber security for space infrastructures will only grow in scope and variety [1]. As most of the world's terrestrial critical infrastructures – communications, financial services, transport, logistics, weather monitoring and a multitude of other services[3] - are intrinsically dependent on space-based assets [2], preventing their compromise is a critical and urgent need.

The current trend is for the space sector to grow even faster in the future [3], with the space capabilities becoming more and more commercially competitive, driven by falling costs of launch and rapid technology developments. Despite the space industry's sophistication and the increasing dependency on the space infrastructures, cyber security issues are somewhat under-recognized by the infrastructure providers and policymakers alike; they tend to lag developments in other high-technology sectors. In this article, we examine the current threat landscape of space infrastructures and security challenges to outline the issues that need to be tackled in the formulation of cyber security solutions for space as well as provide recommendations for developing a policy framework for space security.

---

[1] Global Innovation Chair Professor and Chief Cyber Strategist, The University of Newcastle, Australia (Email: vijay.varadharajan@newcastle.edu.au).
[2] Distinguished Professor & Chair in Cybersecurity, Lancaster University, UK (Email: Neeraj.suri@lancaster.ac.uk)
[3] Space systems provide the means of communicating vital information to keep the power grid synchronized and stock market transactions timed. Should the availability of such timing become impacted, the economy could be crippled, potentially leading to shortages of food, water, medicine, and commodities. Moreover, global navigation services, such as the Global Positioning System (GPS) that are used for aviation, shipping and in daily civilian travel depend on space infrastructures. Satellites provide aerial coverage to view areas struck by natural disasters, enable live reporting of events, and provide information for organizing coordination of international relief efforts; the use of space-based surveillance systems has also been the basis for warfare deterrence.

The Typical Scope of Space Systems

Space systems are assets that either exist in suborbital or outer space or ground control systems, including facilities used in launching these assets [4]. Space systems are usually divided into three technological and operational segments namely the ground segment, the space segment, and the link segment. The space segment comprises groups of satellites in orbit (as well as launch vehicles designed to release satellites into space). A satellite contains a payload, the equipment designed to carry out the satellite's function, and a bus, which houses the payload and remaining satellites systems. The link segment consists of the transmission channels between the satellite and the ground station, as well as between satellites. The ground segment consists of all the ground elements of space systems and allows command and control, and management of space objects such as satellites as well as the data arriving from the payload and delivered to the users. All these segments can be exposed to a range of cyber threats.

Space organizations are organizations that build, operate, maintain, or own space systems. Space systems, however, are somewhat more complex than terrestrial digital infrastructures from a technology development, ownership, and management perspective. Cyber vulnerabilities pose serious risks not just for space-based assets themselves but also for ground-based critical infrastructures. From a technology perspective, it is easy to imagine an attacker attempting to interrupt a nation's commerce by attacking cloud services offered by companies such as Amazon or PayPal or a banking institution. However, nowadays these companies invest heavily in cyber security and are constantly monitoring their systems and networks for malicious activities and vulnerabilities. For an attacker, a simpler and probably a more productive route would be to target space infrastructures that provide connectivity to financial systems and services and attack the space organizations that provide and operate these satellites enabling such services. The ability to impact multiple systems by compromising a central point of failure makes space systems attractive targets. Not only do they offer a vast attack surface but there is also a lack of security regulation governing space systems, despite their critical nature. Often space systems are overlooked as being part of the underlying infrastructure for critical systems and are not subjected to same level of security standards. Furthermore, the situation is exacerbated due to the ambiguity that often exists when it comes to the responsibility for cybersecurity in space and its ongoing management. Moreover, commercial transformation of space capabilities also raises some fundamental questions as to how best to regulate the activities of the commercial actors in space.

## 2. What Makes Space Infrastructures Vulnerable?

There are many operational satellites using older technologies such as prior-generation processors[4], which are susceptible to cyberattacks due to weak security functionalities, if any. For instance, they may have their security credentials hardcoded and there can be insecure communication protocols making them vulnerable to attackers. Furthermore, as more and more commercial actors begin to access space and start offering a range of services, it dramatically increases the attack surface. A summary of different types of threats and vulnerabilities in space systems is given in Table 1.

---

[4] As processors for use in space need to be hardened against radiation and entail long operational lifetimes of 10-20 years, typically simpler and well-tested designs are used to maximize their robustness. Furthermore, the relatively small volumes of space deployed processors also limit the use of cutting-edge technologies that are geared towards performance and economies of scale than longevity.

## Vulnerabilities in the Ground Systems

Compromising the ground station infrastructures is the easiest way to attack space systems, as it provides the software and the hardware required to legitimately control and track space objects using existing terrestrial networks and systems. It is important to note that there is also the user segment of the space infrastructure, which can be thought of as an extension of the ground segment for the end-users of a space-based service. This can itself be a distributed infrastructure providing interfaces to various applications and services that can interact with satellite signals directly or with other ground segment systems.

In the same vein as attacks on enterprise infrastructures, the space attack vectors involve techniques such as exploitation of misconfigurations and software vulnerabilities in systems, gaining unauthorized access to critical services, injection of malware and use of phishing to obtain sensitive credentials. For instance, this could involve exploiting web vulnerabilities or luring the ground station personnel to download malware (such as Trojans) to their devices to take control of the satellites and sabotage them. Data modification attacks can result in corrupted commands to space assets which could lead to either wrong action being taken or even no action taken that could itself result in dangerous situations. Furthermore, physical attacks such as unauthorized access to ground stations and other physical assets can lead to disabling the ground station, directly compromising the operation of the space mission, and taking control of the space assets and their services without technically attacking the systems. In fact, some of the interesting attacks against space systems have been on the services that enable them. For instance, vulnerabilities in the ground systems or in the satellite data receivers can allow the attacker to infiltrate the ground network and to remain there undetected. Another common threat is the introduction of a malware into the satellite's hardware and software systems (e.g., via the supply chain), which can also compromise the ground systems at a later stage. The Turla attack [5] on a satellite Internet provider enabled the attacker to steal IP addresses. This attack was not easily detectable because it was dependent on whether the attacker and the legitimate user were using the IP address simultaneously. It allowed an attacker to inject false data to the user systems connected to that IP address, such as an autonomous drone, leading to its crash. Hence such attacks can remain stealthy and unlikely to be detected by intrusion detection systems. In June 2018, Symantec Corporation reported a sophisticated attack launched from computers in China infiltrated computers in satellite operators, defence contractors, and telecommunications companies in the United States and southeast Asia [6]. It claimed that the attackers infected the computers that controlled the satellites, which could have enabled them to change the positions of the orbiting satellites and disrupt data traffic.

## Vulnerabilities in the Space Segment

The space segment comprises groups of satellites in orbit as well as space stations and launch vehicles designed to release satellites into space. A satellite itself contains a payload and systems designed to carry out the satellite's functions. For instance, these systems are responsible for receiving and processing uplink and downlink signals, validating, decoding, and sending commands to other subsystems, and controlling the stabilization and orientation of the satellite etc.

A spacecraft must be able to communicate, maintain orbit and deliver power to mission-significant components. If the ability to communicate is subverted, for instance, by attacking the Telemetry, Tracking & Command (TT&C) systems in a satellite, then the spacecraft will be unable to support its mission. Similarly, if an attacker targets the satellite's orbital dynamics, e.g., by modifying attitude control function, it could compromise the satellite's mission.

In general, space systems can be subjected to at least four types of cyberattacks. Spacecraft could be vulnerable to command intrusions (giving bad instructions to destroy or manipulate basic controls). There can also be malicious control of payload and attacks such as denial of service (sending too much traffic to overload systems). Malware could also be used to infect systems on the ground (like satellite control centres and user systems), as well as the links between them, spoofing the communications from an untrusted source as a trusted one or replaying, interrupting, or delaying communications.

From an operational perspective, the space environment presents certain unique challenges leading to situations which few consumer hardware systems will encounter. Once the space asset is deployed, developers and operators are unlikely to have direct access to the hardware of the space asset again. This lack of physical access further intensifies the need for cyber security measures. For instance, the physical damage of the ROSAT satellite in 1998, caused by the satellite pointing towards the sun [7], was later revealed to be due to an operational error resulting from the execution of a command outside of the safe operational limits of the satellite. This was thought to be the result of a cyberattack by Russia [7]. An attacker operationally compromising a satellite can permanently wrest control from its legitimate owner, "locking out" the owner and gaining control of the satellite's strategic capabilities. In fact, this amounts to the ultimate transfer of ownership. A satellite which cannot be contacted by its legitimate owner is just as dangerous as a malicious satellite placed on a collision trajectory. Even slow-moving satellites, travelling at 18 000 mph, could be as much a threat as any anti-satellite missile.

Communication Vulnerabilities

The most common threat against the communication channels (uplink and downlink channels) is that of jamming, which compromises the GPS systems [8]. GPS jammers send signals over the same frequency as the GPS device, to override or distort the GPS satellite signals. GPS jammers are widely accessible and cheap to purchase, rendering them available to less sophisticated state and commercial malicious actors. In Nov 2018, Russia was suspected of disrupting GPS signals, when Norway and Finland participated in NATO's Trident Juncture exercise [9]. The attacker can also alter the legitimate signals or even change them completely. This is sometimes referred to as satellite hijacking [7], which amounts to reusing a satellite for another purpose. Broadcast signal intrusion is a form communication hijacking, where broadcast signals of a satellite are hijacked either by overpowering the original signal at the same frequency or directly breaking into the transmitter and replacing the signal. A well-known incident is the Max Headroom Broadcast Signal Intrusion Incident [10], where the attacker generated a more powerful signal over the satellite and land-based microwave links.

GPS spoofing involves the manipulation of the GPS signal and is more dangerous than jamming because it appears to the user that the GPS is working as intended. A system that can execute a software-defined spoof attack is easy to develop with low costs (e.g., $1000 or so to build as demonstrated in [11]). For instance, it is believed that, in Sept 2011, Iranians successfully captured an

American RQ-170 Sentinel drone by reconfiguring the coordinates of the GPS signal to make the drone land in Iran instead of its base in Afghanistan [12].

All it takes is the production of a relatively inexpensive spoofer, and an attacker can command and control the uplink signal to a satellite. If the downlink from a satellite is spoofed, false data can be injected into a target's communications systems, fooling the receiver into calculating an incorrect position. Intentional alteration of data communicated to the spacecraft can have a catastrophic effect, if either no action occurred (e.g., command is discarded) or a wrong action taken by the onboard systems in the spacecraft.

Often the communication channel between the satellite and ground systems involves the use of radio frequency signals sent over the air, which are susceptible to interception. If the traffic is unencrypted, the attacker could also intercept and eavesdrop on the satellite traffic. In the near-term, these kinds of attacks will likely remain, coming not only from nation state actors, but also from well-resourced non-state actors (e.g., criminal groups seeking financial gain), as more communications capabilities come online via space.

Supply Chain Vulnerabilities

Another major issue in space system security arises due to the complexity of supply chain and vendor ecosystem of government funded systems. Usually, the specialized components needed for space assets are not all developed by a single manufacturer. In fact, to keep the costs down, space organizations often purchase components from catalogues of approved vendors around the world. The approval process for these vendors does not necessarily specifically include cyber security vetting standards. When a space organization purchases a component from a vendor, for instance, it has little control over the code written by a software developer of that component. This lack of insight introduces considerable cyber security risk.

In addition to vendors being vulnerable across the system supply chain, space organizations often tend to work with several research institutions, who may have their own vulnerabilities. Naturally collaborations across multiple partners exacerbate potential supply chain security issues, which make it difficult to ascertain who should be operationally (and financially) responsible for the cyber security of a system at various point of the space asset's lifecycle. Hence the security challenge in the space asset supply chain life cycle is caused by the complexity of development, management, use and the ownership of space assets.

Furthermore, cloud infrastructures form an integral part of service provision, and hence are often used for data storage and processing by ground station systems. These cloud systems are owned by other commercial providers, and vulnerabilities and failures in in these infrastructures can have adverse effects, such as hindering operations of satellite real time systems and denial of service attacks on the satellite receivers.

Unlike critical infrastructures, space assets are often not owned by the same organization that manages the infrastructure, which results in uncertainties concerning liability in the event they are attacked; the longer lifespan of the space asset itself complicates these issues even more. Space missions can last decades and because of this, security concerns are exacerbated by unpatched legacy systems. Not dissimilar to industrial control systems, space assets are built to last and because they are functional in the field for long periods and are mission critical, system downtime is not usually an option. This makes space assets difficult, if not impossible, to patch for security flaws, when they are discovered. Furthermore, with the increasing use and connectivity to Internet of Things (IoT devices), attacks on space satellites can cause wide disruption to communication channels endangering national as well as international security [13].

| Ground Segment Cyber Threats | <ul><li>Cyberattacks and intrusions against ground systems and infrastructures supporting space operations as well as user applications and interactions<ul><li>Exploitation of misconfigurations and software vulnerabilities to gain unauthorized access to critical systems and networks</li><li>Injection of malware into systems that control space operations, and into satellite data receivers and transmitters</li><li>Use of phishing to obtain sensitive credentials</li><li>Introduction of malware in ground station system components via supply chain</li></ul></li><li>Physical attacks on ground station infrastructures</li></ul> |
|---|---|
| Space Segment Cyber Threats | <ul><li>Adversary using space situational awareness to target satellites, attack space assets with counterspace weapons, and assess the effectiveness of those attacks.<ul><li>E.g., detecting the presence, location, and identity of satellites to track and target attacks to deny strategic capabilities.</li><li>Target attacks such as using electromagnetic pulse actuators to cause failures in the victim satellites or radio frequency transmitters to cause jamming and spoofing attacks.</li><li>Any technology that enables close approach to inspect and repair satellites could be used to conduct an attack on other satellites, resulting in temporary or permanent damage.</li></ul></li><li>Lock out: Adversary taking control of space assets in an unauthorized manner by operationally infiltrating and compromising satellites and locking out the legitimate satellite owners.<ul><li>A satellite which cannot be contacted by its legitimate owner is just as dangerous as a malicious satellite placed on a collision trajectory</li><li>Even slow-moving satellites, travelling at 18 000 mph could be as much a threat as any anti-satellite missile.</li></ul></li><li>Exploitation of the software and hardware vulnerabilities of the space assets<ul><li>Via command intrusions such as giving bad instructions to destroy or manipulate basic controls</li><li>Via malicious control of payload and attacks such as denial of service attacks to overload systems</li><li>Via introduction of malware in satellite system components via supply chain</li></ul></li><li>Satellite hijacking: Attacker altering the legitimate signals of a satellite and reusing it for another purpose.</li><li>Monitoring and tracking of military and other sensitive activities and locations on the earth using space assets such as surveillance satellites</li><li>Physical attacks against space assets<ul><li>E.g., space-based robotic arm technology for grappling other satellites</li></ul></li></ul> |
| Link Segment Cyber Threats | <ul><li>Adversaries conduct attacks to disrupt, deny, deceive, or degrade space communications and services<ul><li>Via jamming to prevent users and other space assets from receiving intended signals, e.g., uplink jamming and downlink jamming. Uplink jamming is directed toward the satellite and impairs services for all users in the satellite reception area. Downlink jamming has a localized effect because it is directed at ground users.</li><li>Via spoofing to deceive the receiver by introducing a fake signal with erroneous information.</li><li>Attribution of such attacks and distinguishing them from unintentional interferences can be difficult</li></ul></li><li>Eavesdropping on satellite communications<ul><li>Especially if the traffic is not encrypted</li></ul></li></ul> |

Table 1: Summary of Threats and Vulnerabilities in Space Systems

## 3. Security Challenges

Given the rapid growth of the space sector and the increasing ability to manipulate and exploit the vulnerabilities in space systems by an increasing number of diverse space actors, cyber security for space poses several unique challenges. Not only is space becoming increasingly congested, contested, and competitive, but it is also becoming more commercial. The danger with growing space activities and the proliferation of space-capable actors is that it can lead to mistrust among the parties, which can potentially lead to miscalculations and misunderstandings especially with new technologies. Let us now examine some new and unique security challenges in space infrastructures.

- The current state of the art in security in space systems is often based upon strong boundary protection in the ground segment together with encryption to secure communications between the ground station and the space objects. Onboard the space object (such as a satellite), often the assumptions made are that its components are trusted based on the assurances in the supply chain. This, in turn, means that the spacecraft themselves are designed with few if any security defence mechanisms. For instance, if an adversary were able to gain access to the ground segment or insert malware into a spacecraft component, then there are often few or no protections to prevent them from directly controlling the space segment.
- A consequence of lack of built-in security measures in space systems is that it provides a new opportunity for the attackers to discover and exploit vulnerabilities, and maliciously manipulate remote space objects. Scarce documentation and lack of source code availability create the "security through obscurity" mentality with which vendors often develop these space products.
- In terrestrial network systems, we regularly employ intrusion detection and prevention systems (IDS/IPS) to monitor and respond to threats in infrastructures. Similar technology will be required for space systems to observe and tackle potential attacks onboard satellites such as data protocol and RF-based attacks. Intrusion detection and prevention technologies leveraging machine learning to detect and block cyber intrusions onboard space objects would be the natural approach to consider in future space systems. However, this can introduce additional issues related to competing power and memory requirements and scalability, as well as some additional trusted hardware and software, which themselves need to be secured. Furthermore, having an IDS/IPS technology should not act as a replacement for secure design and development of space systems.
- The remoteness and lack of physical access to space assets create some unique challenges. One such challenge arises from the need to perform software updates on space system components, e.g., satellite firmware updates. Unpatched software exposes space systems to attack vectors that are openly documented and available for exploitation. However, these updates can only be performed when the satellites are visible to ground stations and may require more than a single fly-by. Furthermore, a firmware update that may need to be delivered to multiple satellites, by beaming them to a single satellite across multiple passes over a ground station, and then that satellite transmitting them to other satellites requiring the same update. Software updates can introduce vulnerabilities, either inadvertently through a legitimate transmission of the update, or through an attacker using this circumstance to purposefully inject flaws into the space object [14]. For instance, in the case of the space probe Phobos 2 [15], a software update inadvertently caused the spacecraft to lose its lock on the Sun, which drained power and ceased communications. Techniques such as software attestation can enable the software to prove its identity, thereby increasing its trustworthiness.

- Despite the challenges in dealing with remoteness, the software problems afflicting space objects are somewhat similar to those afflicting systems on Earth. These problems can be particularly pronounced in space systems, as security has not been incorporated into the design of space computing systems in the first place. Furthermore, there can be many components in space systems with legacy software, pre-dating the time security was considered important.
- When it comes to detecting malicious behaviour, an important issue is that of intent of an entity's actions. Often, when monitoring manoeuvres of foreign space objects, there is little information beyond what is being perceived with telescopes and radars. These observations might reveal the trajectory of the space object and some physical characteristics, but it can be difficult to determine the nature of a space object's mission without further information. This makes the assessment of intent even more challenging when it comes to the movement of space objects. In the absence of further additional information, there are only the official state policies of others on their space activities to provide the necessary context for what certain actions might mean. Such policy declarations are often general in nature and do not necessarily cover specific classes of activities, which adds uncertainty to the decision making.
- As many strategic military systems (such as missile systems) rely on space infrastructures for navigation and command and control, cyberattacks on space systems would undermine the integrity of strategic weapons systems and have the potential to obfuscate the originator of the attack. As cyber technologies are increasingly within the grasp of non-state actors, they create hitherto unparalleled opportunities for even small malicious groups to instigate high impact attacks. In fact, the asymmetricity in cyber is exacerbated in the space domain, where offence is easier than defence, both technologically as well as geopolitically.

3.1 Specific Security Problems

We will now briefly outline some specific security issues that present significant problems in the design and deployment of secure space systems.

- Secure Positioning: Space situational awareness involves the detection and characterization of a space object, including its location, and the ability to identify it and predict its future location. So, when it comes to space systems, first question is how to securely identify the correct location of a space asset such as a satellite. Secure positioning technologies aim to find the correct position of a device in the presence of an attacker determined upon falsifying it. Today's positioning systems are often vulnerable to location spoofing by which devices can cheat on their own positions or can manipulate the measured positions of other devices [16]. This is not a problem that can be fixed by simple upgrades, as existing positioning systems rely on legacy distance measurement techniques and protocols that were designed without security considerations or with security as an after-thought. There is a clear need to develop a new accurate positioning infrastructure that is secure and private by design. Such an infrastructure should provide resilience to location and time spoofing, location verification as well as support identity and location privacy.
- Friend or Foe: The next follow up question is how to determine the identity of the space asset, to detect a rogue space asset from a benign one. For instance, a rogue satellite can pretend to be a legitimate one when communicating with other satellites or with the ground station. This requires not only positioning and satellite ranging schemes to be secured ensuring that two legitimate entities cannot occupy the same physical space, but also incorporating security measures with authenticated attributes that can reliably verify the identity of the space asset and having secure protocols for communications between satellites and with ground station.

- <u>Lightweight Security Protocols</u>: Primary security protocols in space involve authentication of space assets and the establishment and management of keys being used to secure communications. There are lots of research works in the design of security protocols including lightweight protocols such as in [17,18]. Despite this, as far as we are aware, there is no clear guidelines recommending the usage of specific lightweight security protocols for the space industry. Furthermore, the implementation of security functionalities in protocols always introduces additional overheads in terms of computation and memory. The severity of the additional overheads and whether they are acceptable depends on the risk and the performance constraints associated with the space mission at hand.

- <u>Security Key Management</u>: When it comes to large scale space systems with many constellations, management of keys used in security protocols and communications can pose major challenges. First, there can be a large number of keys that need to be managed, from generation to updating to revocation. Depending on the type of key management schemes being used, there can be different types of keys as well as different structures such as hierarchy of keys and group keys. For instance, there can be a key hierarchy, with the space asset having an entity key, session keys for different sessions with other satellites and ground station as well as data keys protecting data within a session. From a security perspective, these keys need to be updated regularly. How often the updates need to be done is dependent on the type of service or communication that the keys are being used to protect. Higher the frequency of updates greater is the security but requires greater management procedures, and hence greater overhead. When it comes to dynamic key management, the secure provision of keys to new space assets and secure revocation of keys from old ones that need to be removed, can pose significant technical challenges. This will be the case, for instance, when satellites dynamically enter and leave a constellation. Revocation of keys is often more challenging, as it needs to be achieved in such a way that it does not impact the keys of the other entities in the constellation. Furthermore, storage of digital information such as keys can be affected by the harsh operating environment in space. The cosmic and thermal radiation can affect the key state in devices, for instance, leading to key bits getting flipped and altered. Hence, key management systems in space assets must be equipped with fault-tolerant mechanisms to deal with such adverse conditions. Emerging technologies such as quantum computing can offer new key management approaches to securing communications in the future. For instance, quantum entanglement can enable distribution of keys to ground stations at the same time. However, at present, there are still some significant challenges associated with distance and system complexity in quantum technologies that need to be overcome.

- <u>Secure Routing</u>: Routing in space has certain unique characteristics that are different to traditional terrestrial networks such as dynamically changing network topology, long and variable propagation latency, asymmetrical forward and reverse link capacities, high bit error rate, intermittent link connectivity and lack of fixed communication resources. There has been a lot of work on routing protocols in space (e.g. [19]) and more recently on secure routing protocols such as in [20]. As the distances between satellites in different planes can vary with the movement of satellites, the routing algorithms should not only establish an effective path between two objects but also maintain the communication path. There is a clear need for efficient secure routing protocols with the increase in the attack surface due to the openness of space communications. An internal routing attack attempts to capture satellites and reprogram them, whereas in an external attack, the attacker is not an authorized participant of the network. In terms of routing process, during route discovery phase, when the routing tables are being built, the attacker will attempt to disrupt the construction of the routing table so that the network space objects get incorrect information leading to attacks such as false routing attacks and denial-of-service attacks. In the data delivery phase, when the packets are forwarded over the paths as per the routing table, the attacker can compromise the forwarding data via attacks such as eavesdropping, data modification and replay attacks. In the route maintenance phase, where update packets are propagated notifying changes in the network topology, the attacker can modify or stop these packets using attacks such as Sybil, wormhole and blackhole attacks. The work in [21] provides a survey of these attacks in sensor networks which are applicable to the satellite domain.

- <u>Secure System Management</u>: Secure management of services and their attributes is a critical issue when it comes to designing large scale space systems. For instance, the authorities involved in the management of attributes such as keys, routing tables and other parameters used by the various satellite services. Design choices in the development large-scale distributed systems and fault-tolerant systems are relevant here. For instance, there are different approaches using centralized or distributed or decentralized architectures with dynamic on-demand as well as more static and pre-computed mechanisms [22]. Each of these choices has different security and trust implications. A centralized approach offers a single logical trusted authority with more control over the security attributes such as keys and the established routes but creates a single point for attack and hence security failure as well as potentially creating bottleneck and congestion issues. On the other hand, distributed and decentralized approaches can provide improved fault tolerance and greater resilience, while increasing the increase the attack surface and creating challenges in distributed trust management. There is a need to synthesize the relevant knowledge on distributed systems and fault tolerance [23] and develop appropriate security architecture for large scale space systems with multi-layer constellations, explicitly addressing security and trust constraints, and their resilience characteristics. For instance, specification of the scope and jurisdiction of the security management authorities in terms of their coverage of space assets as well as ground segments, and the mechanisms to provide resilience against malicious system failures and attacks. Furthermore, as the authorities managing large constellations will themselves be distributed, secure coordination and handover mechanisms will be required. <u>Trusted Computing Base</u>: At a simplistic level, a space asset such as a satellite can be thought of composed of hardware and software systems that process, store, and transmit data, each of which could potentially serve as an attack surface especially when it is operating in a contested environment such as space. A constellation of satellites can be viewed as a compute cluster comprised of multiple such systems. Hence space assets require security mechanisms to detect when its components are being attacked, for instance, by sensing malicious changes to their internal state (e.g., malware causing unauthorized changes to orientation and altitude) using secure introspective situational awareness mechanisms. The detection of external attacks from surrounding objects can require additional sensors to be installed in the space asset. These security mechanisms can stop the attack (or at the least ignore the effects of the attack) and can even repair themselves on their own. For instance, active defences include the deployment of anti-jamming techniques such as changing the communication frequency to launching a separate jamming or spoofing attack on the attacker to even changing the positioning to reduce the attack impact. In the future, we envisage the space assets to have mechanisms that learn from their past actions and adjust their actions, thereby enhancing their resilience.

There is a need for trusted facilities (a trusted computing base) for hosting these security mechanisms, which themselves must be protected against attacks. This will inevitably introduce additional hardware and software, posing further challenges in terms of power and memory requirements. Further work is required to develop such secure solutions within the footprint of a satellite and achieve the desired effect of cyber resiliency to protect, detect, recover, and respond.

- Securing the Supply Chain: As space assets have continued to grow in complexity, there has been an increasing use of commercial-off-the-shelf components in space systems, enabling malicious actors to introduce untrustworthy vulnerable components in the supply chain. With the increasing involvement of commercial companies, whether they are satellite component suppliers or operators, the risk of not having appropriate supply-chain validation has grown even further. Not only it is paramount that high security standards should be followed to ensure confidence and trust in the supply chain, but also further research work is needed on the design of large-scale secure systems composed of untrusted components.
- User Applications: There are several traditional enterprise security challenges when users interact with space systems, either accessing a service through some software application or directly interacting with the receivers. These include authentication of users and applications as well as secure authorization to ensure that they have the appropriate privileges to carry out the required operations. For instance, authenticating and authorizing dedicated receivers accessing space services or data, as well as applications accessing satellites to reconfigure their services or their payloads. Though such security issues are well understood in an enterprise context, it is necessary that they be applied to systems of systems, consisting of both space systems (such as satellites) and enterprise systems in the ground stations.

## 4. Advocacy: Emerging Technology and Space Trends

New space services are emerging, such as the AWS Ground Station, which is a fully managed service that allows users to control satellite communications, process data, and carry out operations from their desktops and laptops, without requiring the traditional ground station infrastructure (such as from a space agency). This implies that such services can be accessed by users from their desktops or laptops, from anywhere from the world. For instance, using the AWS ground station, the user can download data from satellites and store them in the AWS cloud, and then use applications in the AWS to do processing on the downloaded satellite data. As this gives access to space systems for distributed users from their own devices, there is a critical need to ensure secure access to such emerging space services and the associated operations. For instance, not only users and devices must be authenticated before accessing these services but also there is a need for secure authorization services that control the operations of the users on the space infrastructure and data. Furthermore, security mechanisms are needed to ensure that malicious payload is not uploaded infecting space systems, as well as preventing denial of service attacks.

Another major area of emerging interest is the softwarization and virtualization of space systems and ground station infrastructures. The use of software defined platforms will make space systems more flexible by allowing dynamic configuration of satellite functions to meet changes in demand, thereby helping to improve the efficiency of operations [24]. For instance, a software-defined payload can reconfigure the antenna beam on-demand by sending a new program in uplink communication. This can be used to vary the mission of satellite during its lifetime depending on demand dynamics. Software enabled satellite systems would make satellite systems more adaptable for counteracting jamming attacks by dynamically varying frequencies in jamming areas as well as making them more easily amenable for mobile applications providing coverage to moving targets such as aircraft or vessels or even to cover short temporary events (e.g., natural disasters and exceptional high demand for communication).

Softwarization of space systems is enabled using emerging technologies such as software defined networks (SDN) and network functions virtualization (NFV), providing programmability, flexibility, and modularity that are required to create multiple logical networks, each tailored for a given use case, on top of a common network. SDN and NFV technologies can be applied to both the ground and the space segments of the network infrastructure. Cyber security has a key role to play in these new technologies. For instance, secure smart software enabled satellites can better detect and defend against cyber threats autonomously and update on-board cyber defences to address new threats. They can also diagnose issues with greater precision and back each other up when needed, significantly enhancing resiliency. The virtualization technology with the hypervisor securely containerizing virtual machines helps to optimize memory, on-board processing, and network bandwidth. For instance, it enables the smart satellites to process more data in orbit thereby only transmitting the most critical and relevant information and saving bandwidth costs and reducing the burden on ground station. With the improvements in hardware and software technologies, the deployment such softwarization architectures are becoming more feasible in practice[5]. This will ultimately help to host future data centres and infrastructures in space.

The novelty of software enabled space architectures is that it can provide end-to-end logically isolated network services supporting diverse use cases from multiple tenants, with independent control and management, and which can be created on demand over a common infrastructure. It can also support new network services on-board space platforms, with the capability to provide arbitrary per-flow logic and accommodate rapid topology changes in constellations. However, such softwarization of space will introduce a whole set of new security challenges, as security and trust are critical for the dynamic provision and management of space services and counteracting sophisticated security attacks against space systems [27,28].

---

[5]For instance, such softwarization architectures require SDN Controller, containers for deploying virtual functions, and virtual switches. Depending on the type of satellite functions that are being virtualized, this would require memory of the order of 1 to 2GB with a 64-bit processor and a dependable operating system such as RODOS [25]. Recently, on-board-computer (OBC) technologies for spacecraft -- such as Telos OBC 60 series [26] offering a 64-bit ARM processor, a flash memory of 128GB and a power consumption of 0.2 to 12W -- are becoming available in the marketplace that can cater for these requirements.

Another important technology relevant for space is that of trustworthy autonomous space agents, collaborating with each other to realize overall system goals, carrying out a multitude of tasks, in a dynamic, adversarial, and contested setting. These agents should have the ability to dynamically learn from the environment. As they will be operating under contested environment, they should have mechanisms to protect themselves from attacks from other malicious space objects. They should be capable of making trustworthy decisions under uncertainty and adversarial threats as well as able to adapt to changes in the environment and behave in a goal directed manner involving different levels of forward planning to fulfill their mission [29,30].

Such trustworthy autonomous agents are needed in the establishment of future space facilities such as hosting and managing infrastructures in space stations (e.g., the moon) for further space exploration. They also form part of new generation smart satellites. For instance, such trustworthy autonomous satellites can be used to police routes in space and counteract attacks against space facilities from rogue space entities. The dedicated trustworthy autonomous space entities could even help to constitute new space force for protecting space facilities. This can be seen as a natural extension to the current use of satellites in military conflicts. For instance, several countries (e.g., USA, Russia, and China) have launched numerous small satellites to support military functions over the last decade [31].

## 5. Advocacy on Response and Mitigation: Technological and Policy Solutions

It is clear that mitigating cyber threats in space require both technological as well as policy solutions. Though many of the technology solutions for terrestrial systems can be applicable for space infrastructures, as previously identified in Section 3, space creates certain unique cyber security challenges. Furthermore, as the threat environment is dynamic, the technological solutions also need to be dynamic and adapt to new threat situations. In addition to traditional security mechanisms counteracting attacks such as GPS spoofing and lightweight security protocols, new security architecture and solutions are required to cater for softwarization of space systems, advanced autonomous space agents and managed services enabling user access to space objects, and quantum-based security technologies, as outlined in Section 4. However, a comprehensive approach to effective response and mitigation requires a systematic and unified policy solution that can guide the technology efforts to protect space assets and services. There must be mechanisms for the enforcement of policies, which enable legitimate users and actions while increasing the costs for illegitimate users and their behaviours.

The policy solution needs to address several dimensions as new actors (state, non-state and commercial) and new technologies are expanding and transforming space activities. However, at the present, neither space policy nor cyber security policy is prepared for the challenges created by the meshing of space and cyberspace dramatically increasing the security risks. The commercialization of space with the market incentives to lower costs and entrepreneurial activities such as space tourism and asteroid mining, heighten cyber security concerns. There is also a growing development in the networks of small satellites and new satellite services for use in a range of applications such as agriculture, transportation, and environmental monitoring, producing valuable data, which can be targets for cybercrime and espionage.

The central premise of the policy solution is that it should reflect an end-to-end framework for cyber security, incorporating measures into all stages of space system development and operations. With the increasing reliance of the space sector on commercial technologies and the use of commercial off the shelf components, it is critical that polices should be established to enforce strict cybersecurity requirements for all components of space systems and their supply chains, spanning both civilian and military space assets and activities, for instance, considering the Cybersecurity Maturity Model Certification (CMMC), which has been introduced as a requirement for all defence contractors and providers, including small vendors [32] by the US Dept of Defence. There should also be a supply chain risk management program and software assurance methods within the software supply chain to reduce the likelihood of malware being inserted in components and modules. Enforcing strict cyber security standards in government contracts will help to promote the security of commercial products potentially leading to changes across the whole industry.

Another key concern for the policy framework is the need for appropriate regulations for the commercial space sector. With the growth in the range of space activities the private sector is planning, the regulatory framework would provide commercial space enterprises with regulatory certainty while at the same time allow the states to comply with any of the existing space treaty obligations (such as the Outer Space Treaty [33]). It is critical that private parties are included in the discussions establishing the regulatory framework prioritizing industry led efforts strengthening cyber security and collaboration across different sectors in assessing what is non-negotiable versus acceptable risk. Furthermore, international cooperation and agreement with both traditional and non-traditional allies, including international space supply chain stakeholders, is vital for creating sustainable frameworks for mitigating risk in space in the long-term.

Cyber security skills are an important piece in the overall policy framework. A major challenge in securing space systems is the "systems of systems" aspects, requiring a deep understanding of how such systems work and the various threats and opportunities for the attackers to disrupt them. With space systems, expertise in both systems infrastructures such as servers, networks, and systems as well as knowledge of specialised space infrastructures such as ground control systems and satellites are needed. The policy framework should identify specific steps in developing professionals who have capabilities and expertise in both these areas.

Furthermore, the policy framework should have mechanisms and metrics to identify and assess whether the intended policy impacts are occurring. For instance, these include having mechanisms to measure whether the components being used to develop space systems have the required security capabilities, whether providers of space components follow the security guidelines in developing their products and services, whether there is an increase in the capacity of people with cyber and space skills, as well as whether the policy framework is recognized by the different commercial and state actors, and the policies themselves are explainable and auditable thereby enhancing accountability.